\documentclass[preprint,prb,superscriptaddress,citeautoscript,floatfix]{revtex4-1}

\usepackage{amssymb}
\usepackage{amsmath}
\usepackage{gensymb}
\usepackage{todonotes}
\usepackage{tabularx}
\usepackage{multirow}
\usepackage{bm}
\usepackage{graphicx}
\usepackage[colorlinks]{hyperref}
\usepackage{hyphenat}
\usepackage{colortbl}

\hypersetup{
 colorlinks=true,
 citecolor=blue,
 linkcolor=blue,
 urlcolor=blue}

\begin{document}
\title{Tuning the structure of Skyrmion lattice system Cu$_2$OSeO$_3$ under pressure}

\author{Srishti Pal}
\affiliation{Department of Physics, Indian Institute of Science, Bengaluru, 560012, India}

\author{Pallavi Malavi}
\affiliation{Department of Physics, Indian Institute of Science, Bengaluru, 560012, India}

\author{Shashank Chaturvedi}

\affiliation{Chemistry and Physics of Materials Unit, School of Advanced Materials, Jawaharlal Nehru Centre for Advanced Scientific Research, Bengaluru, 560064, India}
\affiliation{Theoretical Sciences Unit, School of Advanced Materials, Jawaharlal Nehru Centre for Advanced Scientific Research, Bengaluru, 560064, India}

\author{Subhadip Das}
\affiliation{Department of Physics, Indian Institute of Science, Bengaluru, 560012, India}

\author{S. Karmakar}
\affiliation{High Pressure and Synchrotron Radiation Physics Division, Bhabha Atomic Research Centre, Trombay, Mumbai 400085, India}

\author{D. V. S. Muthu}
\affiliation{Department of Physics, Indian Institute of Science,  Bengaluru, 560012, India}

\author{Umesh V. Waghmare}
\affiliation{Theoretical Sciences Unit, School of Advanced Materials, Jawaharlal Nehru Centre for Advanced Scientific Research, Bengaluru, 560064, India}

\author{A. K. Sood}
\email[E-mail:~]{asood@iisc.ac.in}
\affiliation{Department of Physics, Indian Institute of Science,  Bengaluru, 560012, India}

\date{\today}

\begin{abstract}
The insulating ferrimagnet Cu$_2$OSeO$_3$ shows a rich variety of phases such as skyrmion lattice and helical magnetism controlled by interplay of different exchange interactions which can be tuned by external pressure. In this work we have investigated pressure-induced phase transitions at room temperature using synchrotron based x- ray diffraction and Raman scattering measurements. With first-principles theoretical analysis, we show that spin-spin exchange couplings in the ambient cubic phase are affected notably by hydrostatic pressure. The ambient cubic phase transforms to a monoclinic phase above 7 GPa and then to the triclinic phase above 11 GPa. Emergence of new phonon modes in the Raman spectra confirms these structural phase transitions. Notably, upon decompression, the crystal undergoes transition to a new monoclinic structure. Atomic coordinates have been refined in the low pressure cubic phase to capture the Cu-tetrahedra evolution responsible for the earlier reported magnetic behavior under pressure. Our experiments will motivate further studies of its emergent magnetic behavior under pressure.

\end{abstract}

\maketitle

\section{Introduction}

Cu$_2$OSeO$_3$ belongs to an interesting family of chiral, non-centrosymmetric B20 magnetic systems that host a unique magnetic phase diagram consisting of helical, conical and skyrmion lattice structures~\cite{Seki2012,Ruff2015}. Among mostly intermetallic systems like MnSi~\cite{Mulbauer2009,Ishikawa1984}, Fe$_{1-x}$Co$_x$Si~\cite{Grigoriev2007} and FeGe~\cite{Uchida2008}, Cu$_2$OSeO$_3$ is the first insulating material of this family that has been intensively studied for the rich physics of its skyrmion lattice  phase~\cite{Seki2012}. Skyrmion is a few nanometer sized particle-like excitation emerging due to correlated spins in a vortex-like configuration~\cite{Rossler2006,Nagaosa2013} and has been realized experimentally using Lorentz transmission electron microscopy~\cite{Yu2010}, reciprocal space imaging by small angle neutron scattering~\cite{Grigoriev2009}, spin-resolved scanning tunneling microscopy~\cite{Heinze2011} as well as theoretical studies~\cite{Yu2011}.

Bulk Cu$_2$OSeO$_3$ crystallizes in the same P2$_1$3 cubic structure as the other B20 materials with a unit cell containing 8 formula units~\cite{Meunier}. The crystal is comprised of corner-sharing distorted Cu$_4$-tetrahedra (along the body diagonal) of two crystallographically distinct Cu$^{2+}$ ion sites: Cu1 at 4a and Cu2 at 12b with Cu1:Cu2 ratio of 1:3 serving as the backbone of magnetism in the system. It has been shown theoretically that the ground state wavefunction is highly entangled and cannot be factorized into individual spin 1/2 sites~\cite{Judit}. The lack of inversion symmetry of the cubic B20 crystal structure results in large Dzyaloshinskii-Moriya exchange interaction (\emph{D}) and together with the Heisenberg exchange (\emph{J}), it results in the development of helical magnetic ordering. Cu$_2$OSeO$_3$ possess a helical spin ground state below \emph{T$_C$} = 58.8K at zero magnetic field with a fixed pitch of $\sim$50nm~\cite{Seki2012}. This helical ground state further develops into a skyrmion lattice phase on applying moderate external magnetic field followed by a conical spin texture (\emph{B}$>$\emph{B$_{C1}$}) and finally to the field-polarized ferrimagnetic order at much higher field values (\emph{B}$>$\emph{B$_{C2}$})~\cite{Adams2012}. Skyrmion lattice phase is a narrow pocket in the temperature- magnetic field phase space and its formation, size and stability is controlled by different magnetic exchange interactions such as Heisenberg exchange, Dzyaloshinskii-Moriya exchange and magneto crystalline anisotropy. The strengths of these interactions mainly depend on the interatomic bond parameters which can be tuned by chemical doping, disorder or pressure. The metallic members of B20 family MnSi, MnGe, FeGe and Fe$_{1-x}$Co$_x$Si have been well studied for the pressure induced suppression of their ordering temperature \emph{T$_C$} which tends to absolute zero at the critical pressures of 1.5, 23, 18.8 and 7-12 GPa, respectively, above which a non-Fermi liquid (NFL) type dependence of resistivity is observed~\cite{Pfleiderer2007,Martin2016,Barla2015,Forthaus2011}. However, none of the above mentioned behavior is associated with any structural deformation since the cubic symmetry of these B20 chiral magnets remain intact up to 30 GPa~\cite{Forthaus2011,Guo2018,Valkovskiy2007,Wilhelm2007}. In contrast to the above itinerant magnets, \emph{T$_C$} of the insulating Cu$_2$OSeO$_3$ increases with pressure at the rate 0.3K/kbar~\cite{Sidorov} and this contrasting behavior has been attributed to the difference in the nature of the magnetic moments in these systems. Interestingly, hydrostatic pressure is found to expand the size of the skyrmion pocket in \emph{T-H} phase diagram of Cu$_2$OSeO$_3$~\cite{Levatic}. This is due to the interplay of complex magnetic interactions that modify the exchange interaction paths in the Cu$_2$OSeO$_3$ lattice. Hence, it becomes important to study the structural stability of Cu$_2$OSeO$_3$ under pressure in order to understand its interesting magnetic properties. Recent high pressure study by Deng et al.~\cite{Deng} shows substantial enhancement of the skyrmion pocket of Cu$_2$OSeO$_3$ reaching a vast range of 5-300 K with the upper and lower limits being achieved at pressures 7.9 and 26.2 GPa, respectively. The authors also showed that these magnetic transitions in Cu$_2$OSeO$_3$ are associated with a series of structural modulations of the cubic symmetry through orthorhombic, monoclinic and triclinic phases. However, the detailed structural correlation to magnetic properties is still lacking. Our study is focused on structural and vibrational evolution of Cu$_2$OSeO$_3$ under pressure which is important in order to understand the interesting magnetic properties. The detailed structural evolution will provide valuable information for a modeling of the magnetic couplings in these systems under pressure.

\section{Experimental Details}

Polycrystalline samples of Cu$_2$OSeO$_3$ were prepared by standard solid-state reaction~\cite{Bos2008}. A stoichiometric mixture of high purity CuO and SeO$_2$ powders was processed into a pellet, sealed in an evacuated quartz tube and heated to 600$^{\circ}$C for 12h. The process was repeated with intermediate grinding to obtain single phase high purity single crystals of size $\sim$50-500 micron.

Crystals of Cu$_2$OSeO$_3$ were finely powdered and loaded inside Mao Bell type and symmetric diamond anvil cells (DAC) for Raman and XRD measurements, respectively. Both the DACs had two 16-facet brilliant cut diamonds with $\sim$600 $\mu$m culet diameter. 4:1 methanol-ethanol mixture with a freezing pressure of $\sim$10.4 GPa~\cite{Piermarini} was used to transmit the pressure to the sample placed inside the stainless-steel gasket hole of $\sim$200 $\mu$m diameter. Ruby fluorescence was used to calibrate the applied pressure~\cite{Mao1986}.

Pressure evolution of Cu$_2$OSeO$_3$ crystal structure was carried out at Elettra, Italy
using the Xpress beamline ($\lambda$ = 0.4957 {\AA}) at room temperature. Data was collected using MAR 345 image plate. Standard LaB$_6$ crystal was used to calibrate sample to detector distance and orientation angles of the detector. The selected area 2D diffraction pattern was processed using Fit2D software~\cite{Hammersley1998} for conversion into intensity vs 2theta plot. The raw data was refined and fitted using standard Rietveld refinement procedure for the low pressure cubic phase (up to 7.3 GPa) and the decompression data at 3.4 GPa and using LeBail method for the rest of the pressure range in GSAS software package~\cite{Larson2000}.

The unpolarized Raman spectra at room temperature were recorded in a backscattering geometry using Horriba LabRAM HR Evolution Spectrometer equipped with a thermoelectric cooled charge coupled device (CCD) (HORIBA Jobin Yvon, SYNCERITY 1024 X 256). The spectra were recorded using 532 nm DPSS laser illuminating the sample with $\sim$1.5 mW power.

\section{Results and Discussions}
\subsection{X-ray Diffraction}

\begin{figure}[h!]
\includegraphics[width=75mm,clip]{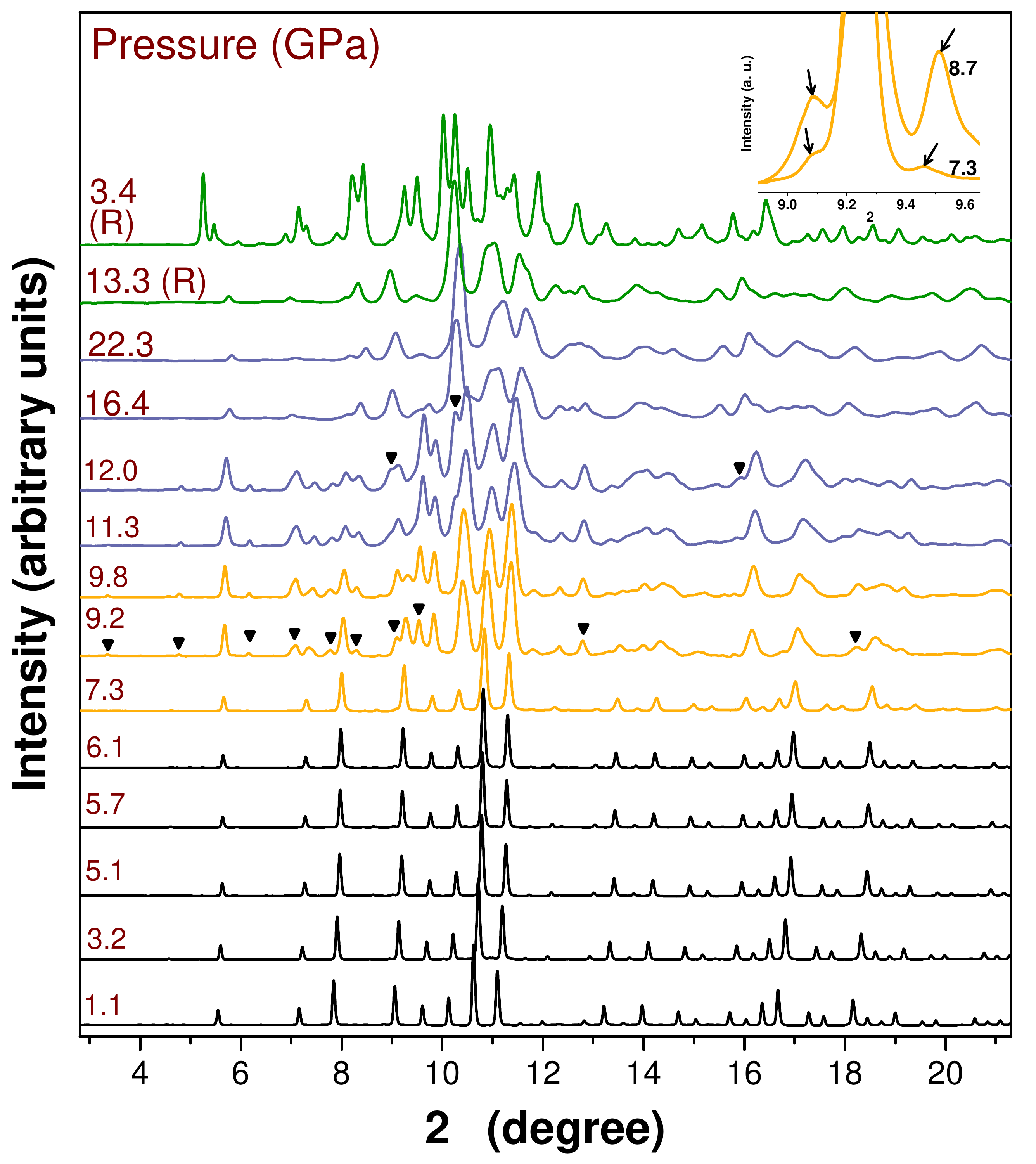}
\caption{\small{Angle dispersive X-ray diffraction patterns of Cu$_2$OSeO$_3$ at selected pressures ranging from 1.1 to 22.3 GPa (the top most pattern is after depressurizing to 3.4 GPa). Arrows indicate the appearance of new peaks. The onset of the first structural transition at 7.3 GPa is indicated in the inset.}}
\label{fig1}
\end{figure}

Angle dispersive powder XRD patterns of Cu$_2$OSeO$_3$ at varying pressure values at room temperature are shown in Fig. \ref{fig1}. The ambient cubic phase with \emph{P2$_1$3} (SG:198, z=8) space group shows stability up to $\sim$7 GPa above which new Bragg reflections emerge in the diffraction pattern ($\sim$3.3$^{\circ}$, 4.8$^{\circ}$, 6.1$^{\circ}$, 7.1$^{\circ}$, 7.8$^{\circ}$, 8.3$^{\circ}$, 9.1$^{\circ}$, 9.5$^{\circ}$, 12.8$^{\circ}$, 18.2$^{\circ}$ and more). Appearance of new Bragg peaks over the existing ones suggests the onset of a pressure-induced first order structural transition. The new phase has been successfully indexed to be monoclinic with space group \emph{P12$_1$1} (SG:4, z=8) that coexists with cubic phase up to $\sim$9 GPa. The transition completes at $\sim$10 GPa followed by another structural transformation around 11 GPa. The phase above 11 GPa has been indexed to have triclinic symmetry with space group \emph{P1} (SG:1, z=8) and is found to be stable up to 22.3 GPa (the highest pressure achievable in our XRD experiment). Our results contrast with the high-pressure XRD measurements up to 10.47 GPa by Deng et al.~\cite{Deng} showing occurrence of intermediate orthorhombic phase between 5 to 7 GPa. The XRD patterns shown in Fig.~\ref{fig1} clearly demonstrate the robustness of the cubic phase in the pressure range of 5 to 7 GPa and hence substantiates the absence of any intermediate orthorhombic phase. The structural transitions in Cu$_2$OSeO$_3$ are found to be path-dependent as indicated by the two top most patterns of Fig. \ref{fig1}. The irreversibility of the structural transitions of Cu$_2$OSeO$_3$ has recently been observed in the high-pressure Raman measurements~\cite{Deng}, though proper identification of the new phase after decompressing back to ambient was not explored. Fig. \ref{fig1} shows that the high-pressure triclinic phase is stable upon decompression down to 13.3 GPa below which the system adapts a different structural transformation channel achieving a metastable monoclinic phase with space group \emph{P12$_1$/c1} (SG:14, z=4) (earlier reported by Effenberger et al.\cite{Effenberger} to be a polymorph of ambient Cu$_2$OSeO$_3$) with a small fraction (17\% in weight) of the cubic \emph{P2$_1$3} one.

\begin{figure}[h!]
\includegraphics[width=75mm,clip]{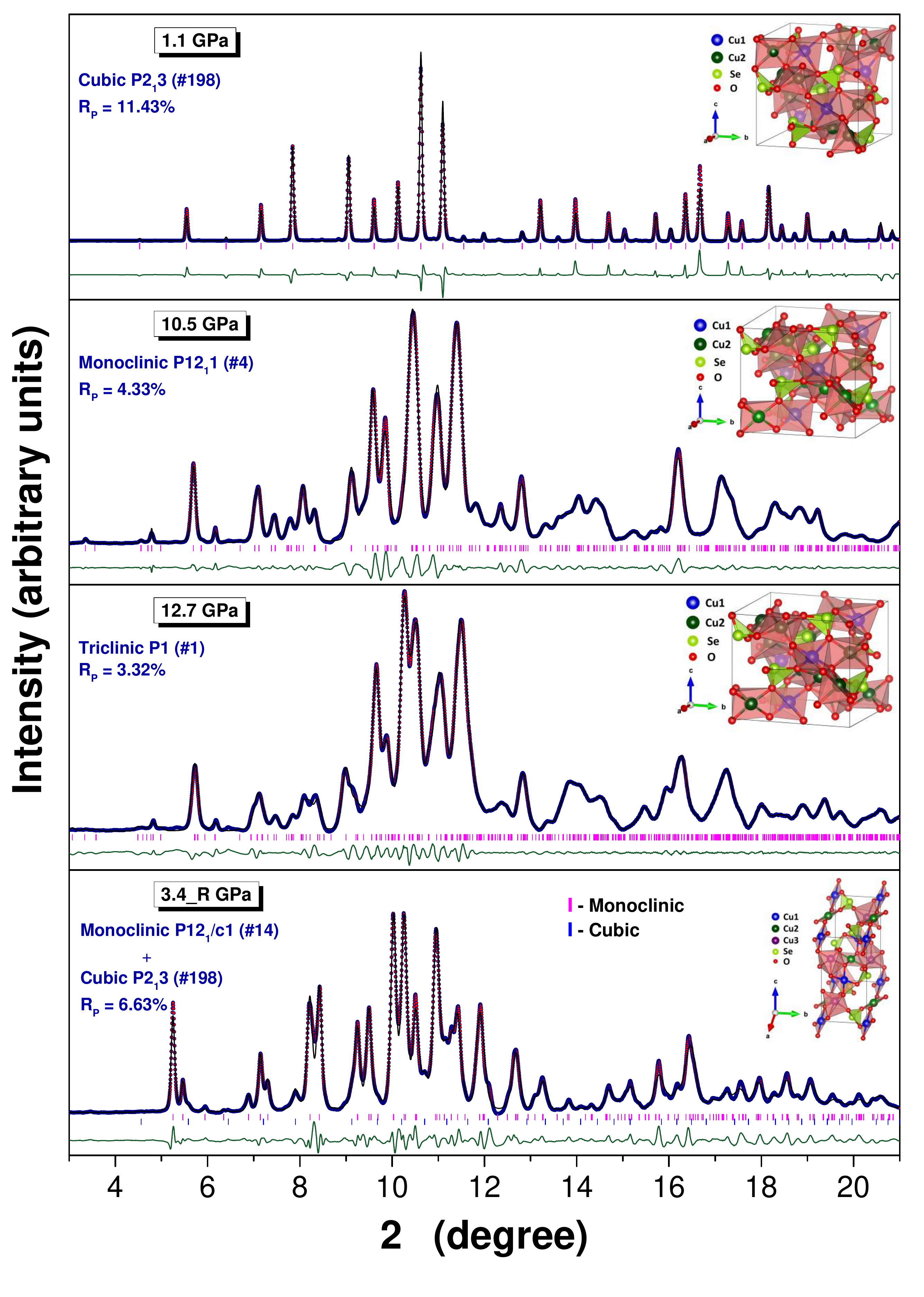}
\caption{\small{Fitted XRD patterns at selected pressure values with the unit cells containing the atoms shown in the insets. Experimental data are indicated by solid circles. Calculated patterns are drawn as black solid lines. Reflection positions are indicated by vertical bars. Lower dark green curves are the weighted differences between observed and calculated profile.}}
\label{fig2}
\end{figure}

Considering the arrangements of Cu-polyhedra in the unit cell, this pressure-released monoclinic structure is in sharp contrast to the cubic polymorph as well as the high pressure monoclinic and triclinic phases which are derivatives of the ambient cubic structure with increased lattice distortion but similar polyhedral environment. While the cubic structure has two types of distorted CuO$_5$ polyhedra viz. trigonal bipyramidal around Cu1 and square pyramidal around Cu2, the monoclinic polymorph contains distorted square planar CuO$_4$ around Cu1 and Cu2 (at Wyckoff sites 2b and 2a) and distorted CuO$_6$ octahedra around Cu3 (at Wyckoff site 4e). The different Cu-Cu distances and Cu-O-Cu angles in the polymorphs are indicative of variation in the magnetic exchange interaction and so the magnetic ordering in these two polymorphic phases. Fig. \ref{fig2} shows fitted patterns at 1.1, 10.5, 12.7 and 3.4 (return) GPa using cubic \emph{P2$_1$3}, monoclinic \emph{P12$_1$1}, triclinic \emph{P1} and mixture of cubic \emph{P2$_1$3} and monoclinic \emph{P12$_1$/c1} respectively. The low R$_P$ values confirm goodness of fit using the unit cells mentioned in the insets of Fig. \ref{fig2}. The refined lattice parameters for the different phases are listed in Table \ref{table1}.

\begin{figure}[h!]
\includegraphics[width=75mm,clip]{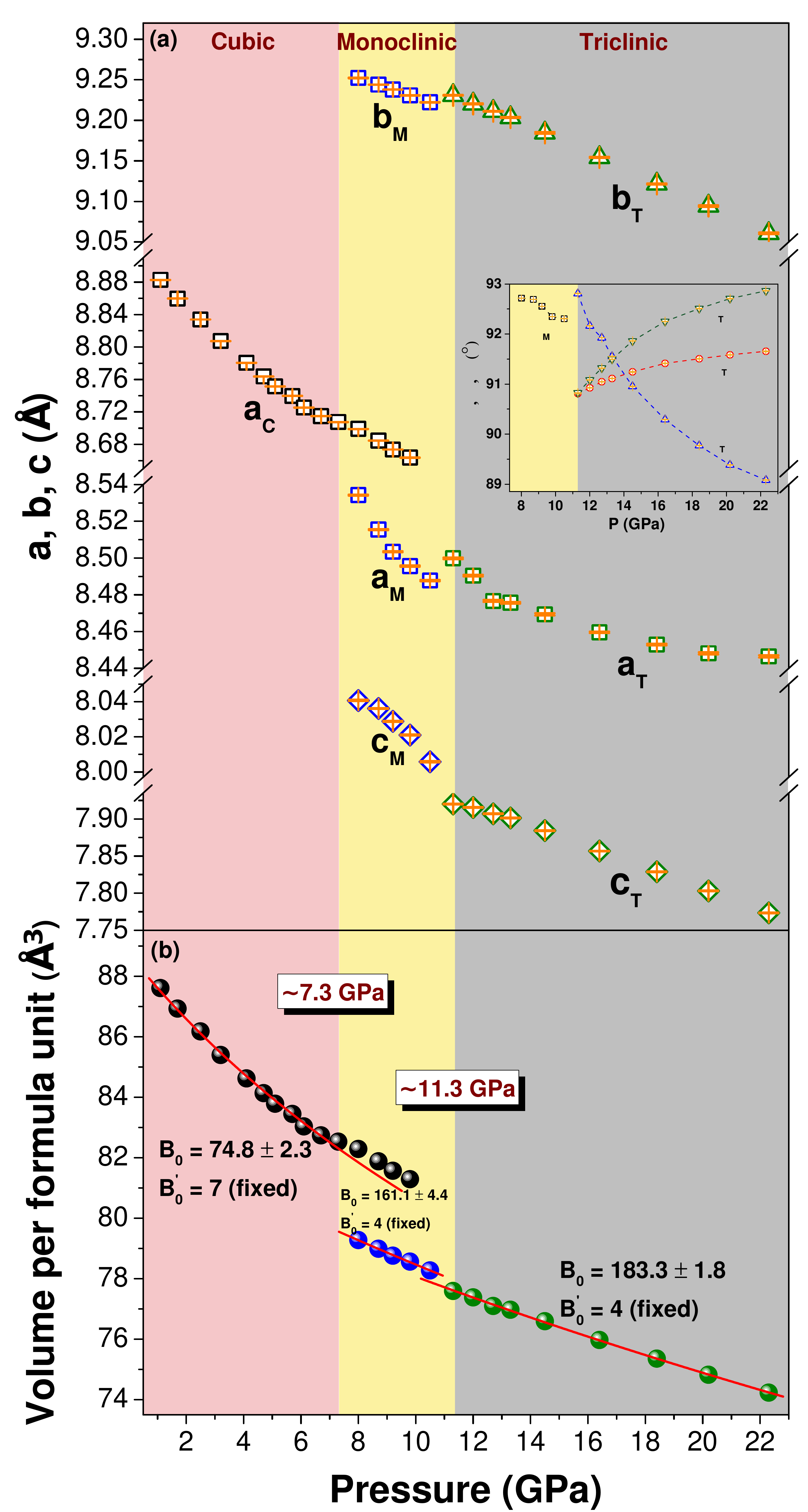}
\caption{\small{(a) Pressure dependence of lattice parameters in cubic (shaded pink), mixed phase (up to 10 GPa) of cubic and monoclinic (shaded yellow) and triclinic (shaded grey) phases of Cu$_2$OSeO$_3$, (b) fitted (red solid line) P-V diagram using 3$^{rd}$ order BM equation of state.}}
\label{fig3}
\end{figure}

The pressure variation of the lattice parameters in different structural phases of Cu$_2$OSeO$_3$ is shown in Fig. \ref{fig3}(a) with the monoclinic and triclinic angles in the inset. The different axes in the monoclinic and triclinic phase show different compressional behavior addressing the anisotropic nature of these symmetry reduced crystal systems. The rapidly falling triclinic $\beta_T$ compared to the monoclinic $\beta_M$ and the contrasting increments in $\alpha_T$ and $\gamma_T$ manifest the increasing disorder in the high pressure triclinic phase. In Fig. \ref{fig3}(b), the volume of the unit cell per formula unit is plotted against pressure and the data in different ranges are fitted with third order Birch-Murnaghan equation of state (EOS)~\cite{Birch1978}. The finite volume discontinuities across the transition pressures of $\sim$7 and $\sim$11 GPa in the \emph{P-V} diagram confirm the first order nature of these structural transitions. The volume data of cubic phase up to 7.3 GPa (before onset of the mixed phase) is used for fitting, giving values of zero-pressure volume \emph{V$_0$} = 88.7 $\pm$ 0.1 $\AA^3$, bulk modulus \emph{B$_0$} = 74.8 $\pm$ 2.3 GPa with its pressure-derivative \emph{B$_0^{'}$} fixed at 7. The values of \emph{V$_0$} and \emph{B$_0$} for the monoclinic phase are 82.9 $\pm$ 0.1 $\AA^3$, 161.1 $\pm$ 4.4 GPa and for the triclinic phase are 82.0 $\pm$ 0.05 $\AA^3$, 183.3 $\pm$ 1.8 GPa, with fixed \emph{B$_0^{'}$} = 4. Increasing values of bulk modulus in successive phases indicate pressure hardening of the crystal.

\begin{figure}
\includegraphics[width=75mm,clip]{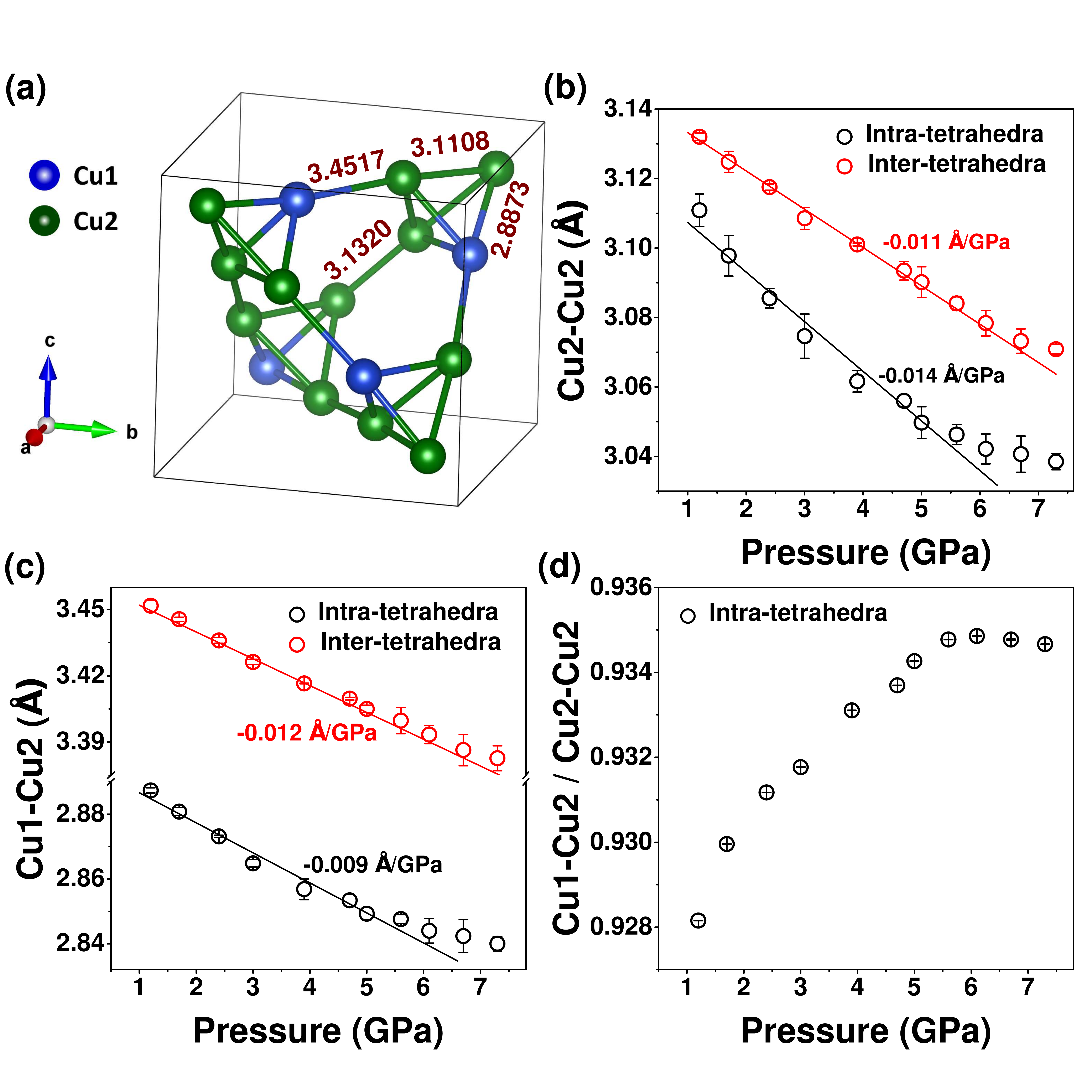}
\caption{\small{(a) Unit cell of Cu$_2$OSeO$_3$ containing the Cu-tetrahedra, (b) Cu2-Cu2 and (c) Cu1-Cu2 distances, (d) intra (strong)-tetrahedron Cu1-Cu2/Cu2-Cu2 ratio as a function of pressure in the cubic phase of Cu$_2$OSeO$_3$.}}
\label{fig4}
\end{figure}

Pressure evolution of the distorted copper tetrahedra of Cu$_2$OSeO$_3$ (Fig. \ref{fig4}(a)) in the cubic phase is illustrated in Fig. \ref{fig4}(b)-(d) in terms of inter-atomic distances of Cu1 and Cu2. Both the intra- and inter-tetrahedral Cu-Cu distances decrease monotonically with pressure but with different rates, indicating increasing anisotropy in the system. Fig. \ref{fig4}(d) represents the quantitative increment in anisotropy in terms of the increasing ratio of Cu1-Cu2 and Cu2-Cu2 distances as a function of pressure. While the dominating super-exchange interaction between Cu1 and Cu2 is attributed to the strong D-M interaction giving rise to exceptionally large $\mid$\emph{D/J}$\mid$ value of 1.95, ferromagnetic Heisenberg exchange prevails between the Cu2 ions ($\mid$\emph{D/J}$\mid$ $\approx$ 0.39)~\cite{Yang}. The size of the skyrmion pocket of Cu$_2$OSeO$_3$ increases with the parameter \emph{JK/aD$^2$} (where \emph{a} is inter-atomic distance, \emph{K} is anisotropy)~\cite{Levatic} , whereas the value of the helimagnetic transition temperature \emph{T$_C$} is directly proportional to \emph{J}~\cite{Janson}. In the next section, with the help of extensive density functional theretical calculations, we will show how the skyrmion pocket and the \emph{T$_C$} evolve with increasing pressure. After transition to the monoclinic phase, anisotropy \emph{K} in the system further increases and the Cu$_4$ tetrahedra becomes more distorted generating several unequal intra- and inter-tetrahedral Cu2-Cu2 and Cu1-Cu2 distances and thus detailed microscopic calculations based on our observed structural evolution are needed to explore the nature of the magnetic interactions in the high pressure monoclinic and triclinic phases of the material.

\subsection{Theoretical Analysis}

\begin{figure}[h!]
\includegraphics[width=75mm,clip]{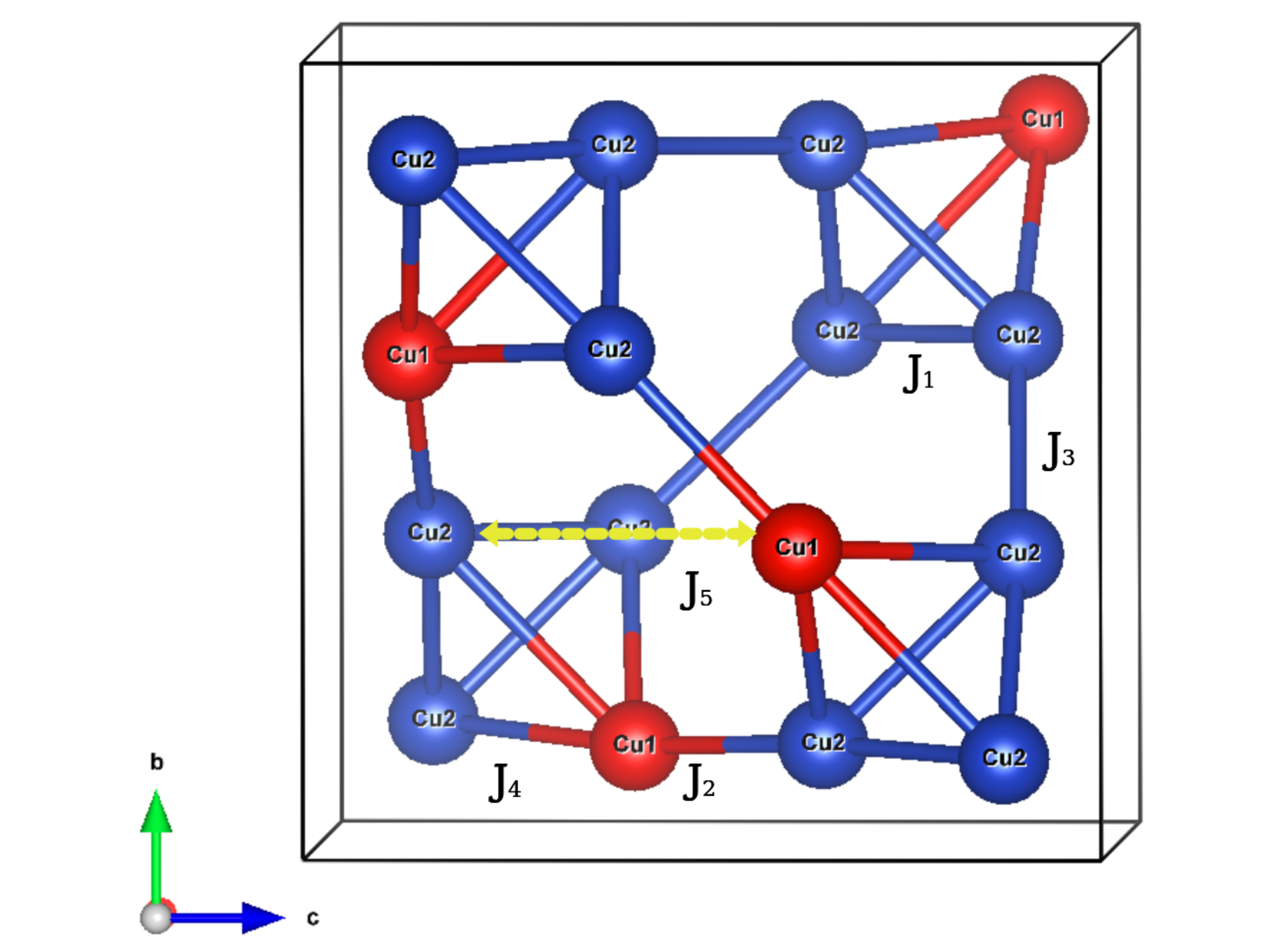}
\caption{\small{(a) Unit cell of Cu$_2$OSeO$_3$ (only Cu atoms are shown) with two types of copper atoms, Cu1 atoms with down-spin (red atoms) and Cu2 atoms with up-spins (blue atoms).~$J_1$ and $J_3$ are ferromagnetic couplings, while $J_2$, $J_4$ and $J_5$ are antiferromagnetic exchange couplings.~$J_5$ is the super-superexchange interaction.~$J_1$ and $J_4$ are intra-tetrahedral and $J_2$ and $J_3$ are inter-tetrahedral couplings.}}
\label{fig5}
\end{figure}

We now present results of first-principles density functional theory (DFT) calculations to estimate various spin-coupling parameters employing VASP package~\cite{vaspJ, vaspG}. Hubbard parameter (U = 7.5 eV) was used to include electron correlations on Cu sites along with $J$ parameter (0.98 eV), based on the method described by Liechtenstein et al~\cite{hubbard}. A generalized gradient approximation~\cite{gga} (GGA) of the electron exchange-correlation energy and projector augumented wave potentials~\cite{paw} were used in our calculations. Plane-wave cut-off energy was set to 500 eV. To determine Dzyaloshinskii–Moriya vector \textbf{$(D_{ij})$}, we performed fully-relativistic calculations with spin-orbit coupling (SOC) to determine total energies of various non-collinear spin configurations as proposed by Xiang et al.~\cite{xiang} To understand the nature of skyrmions, we estimate the symmetric exchange $(J_{ij})$ and antisymmetric exchange $(D_{ij})$ parameters. $J$-parameterized interactions give relative stability of collinear magnetic configurations with parallel and anti-parallel alignment of spins, while the antisymmetric exchange parameter $D$ stabilizes spin-canting. Calculated lattice parameter of Cu$_2$OSeO$_3$ is 9.01 \AA, which is 0.9 $\%$ overestimated with respect to experimental value of 8.925 \AA~\cite{Effenberger}, well within the typical DFT errors. The different inter- and intra-tetrahedral Heisenberg couplings of the Cu$_2$OSeO$_3$ unit cell are shown in Fig.~\ref{fig5}. Our estimates of $J$ and $D$ parameters at 0 GPa are in agreement with earlier work (Table.~\ref{table2})~\cite{Yang}. Pressure does not have same effect on all $J$-coupling constants (Table.~\ref{table3}). $J_3$ coupling weakens with pressure, which corresponds to inter-tetrahedral ferromagnetic coupling between Cu2 atoms, while other $J$ interactions are enhanced with pressure. $J_5$, the super-superexchange interaction exhibits a weak increase from 0 to 6 GPa. Our calculated value of $|D_{4}/J_{4}|$ at 0 GPa is 1.86, which is close to the value of 1.95 reported earlier\cite{Yang}. Application of pressure results in reduction of $|D_{4}/J_{4}|$ to nearly 1.3 at 6 GPa, concluding that intra-tetrahedral $J_{4}$  antiferromagnetic coupling strengthens while $|D_4|$ weakens. $D_4$ interactions are reported to have the highest value of $|D/J|$ in Cu$_2$OSeO$_3$ \cite{Yang} and hence we estimated the value of $J/D^2$ for $D_4$ and $J_4$ interactions. The value of $J_{4}/{D_{4}}^2$ increases from 0.14 at 0 GPa to 0.20 at 6 GPa which contributes to enhance the \emph{JK/aD$^2$} parameter together with an increased anisotropy and decreased Cu-Cu distances and in turn increases the skyrmion pocket size. Also, the Heisenberg exchange interactions $J_1$, $J_2$, $J_4$ and $J_5$ increases with increasing pressure (Table.~\ref{table3}) causing the observed rise in the helimagnetic transition temperature \emph{T$_C$} in earlier studies~\cite{Sidorov}.

\subsection{Pressure dependence of Raman vibrational modes}

\begin{figure}[h!]
\includegraphics[width=75mm,clip]{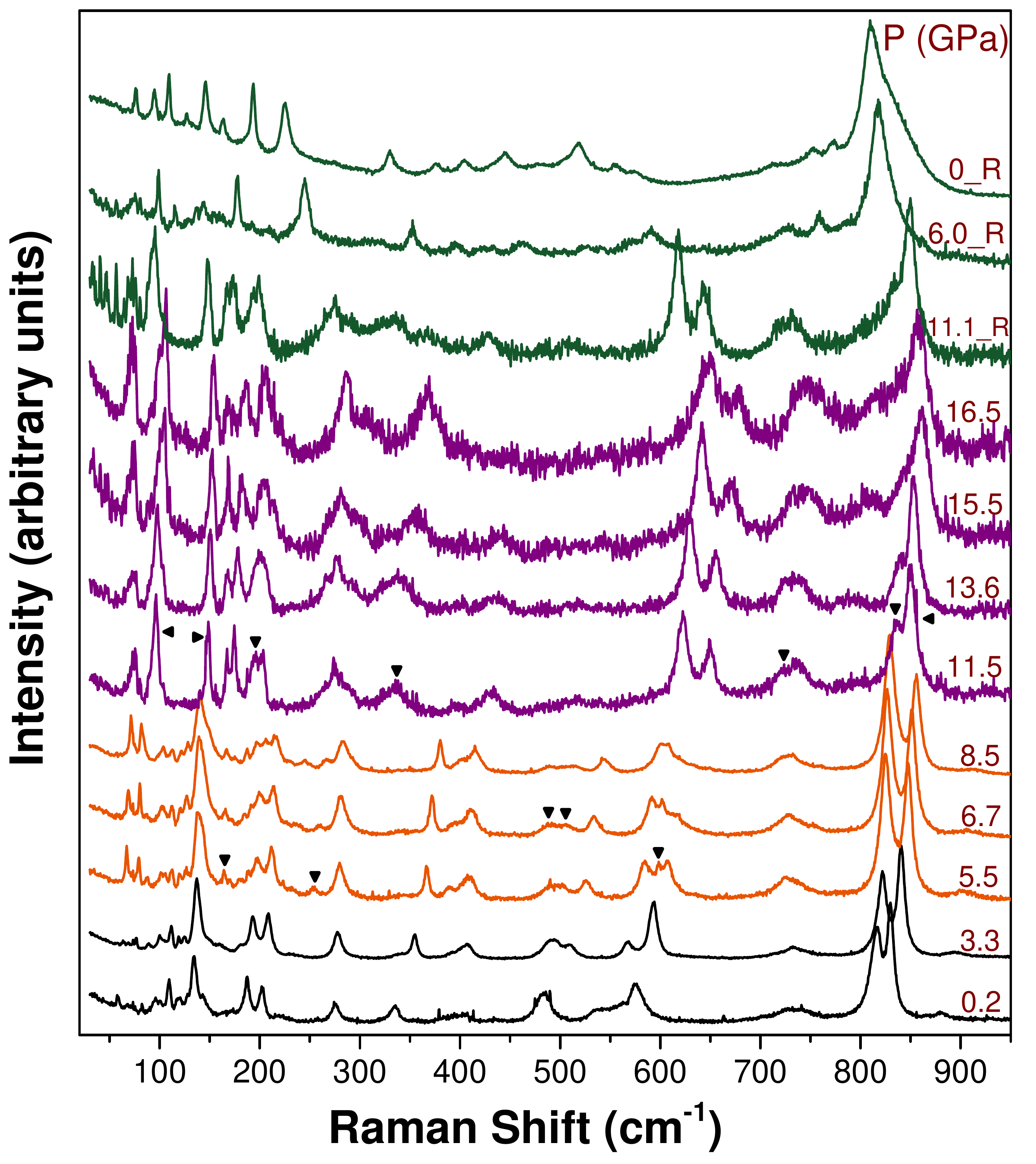}
\caption{\small{ Stacked Raman spectra of Cu$_2$OSeO$_3$ during pressurizing from 0.2 to 16.5 GPa (the top most pattern is after depressurizing to ambient). Arrows indicate emergence of new peaks at the onset of the monoclinic and triclinic phases.}}
\label{fig6}
\end{figure}

Factor group analysis of cubic (\emph{P2$_1$3}) Cu$_2$OSeO$_3$ yields 84 Raman active phonon modes\cite{Gnezdilov} among which 27 modes could be detected in our ambient Raman spectra in the frequency range 50-1300 cm$^{-1}$. Following the mode assignment by Miller et al.~\cite{Miller} and Kurnosov et al.~\cite{Kurnosov} to the Raman and infrared active phonons, the Raman spectra of Cu$_2$OSeO$_3$ can be divided into three distinct ranges of frequencies. While the low frequency region (90-420 cm$^{-1}$) corresponds to the motion of the CuO$_5$ polyhedral entity, the modes in the frequency range 450-600 cm$^{-1}$ can be attributed to the general motion of the oxygen atoms. The modes at frequency higher than 700 cm$^{-1}$ bear vibrational fingerprints of SeO$_3$ units. Fig. \ref{fig6} depicts the effect of increasing pressure on the room temperature Raman signal of Cu$_2$OSeO$_3$ in the range 30-950 cm$^{-1}$.

The highest pressure achieved in our high-pressure Raman measurements was 16.5 GPa. The ambient Raman spectra remained stable up to $\sim$5 GPa above which significant changes started to appear in the scattering profile. The peak intensity of the 546 cm$^{-1}$ mode enhanced significantly along with disappearance of the 143 cm$^{-1}$ mode and emergence of three new phonons at $\sim$165, 254 and 598 cm$^{-1}$ around 5.5 GPa followed by splitting of the 496 cm$^{-1}$ mode into 489 and 506 cm$^{-1}$ around 6.7 GPa. Appearance of new phonon modes in the Raman spectra confirms symmetry lowering (cubic (\emph{P2$_1$3}) to monoclinic (\emph{P12$_1$1})) as established from our high-pressure XRD results. Structural evolution of the cubic phase into monoclinic structure induces significant deformation in the CuO$_5$ polyhedral units causing development of new modes along with vanishing of the old ones in the low frequency band ($<$ 420 cm$^{-1}$). Also, the distortion in the Cu-O bond lengths and Cu-O-Cu bond angles rearranges the vibrational spectrum of the oxygen atoms as reflected in the increasing intensity of the 546 cm$^{-1}$ mode and splitting of the 496 cm$^{-1}$ mode. The onset pressure for the transition is slightly lower for Raman measurements than that of XRD due to higher sensitivity of Raman scattering to probe any deformation of the crystal. Our Raman results also confirm the 11 GPa monoclinic (\emph{P12$_1$1}) to triclinic (\emph{P1}) transition with appearance of new Raman modes at $\sim$95, 138, 197, 334, 716, 786, 820 and 844 cm$^{-1}$ as well as disappearance of some of the existing modes around 11.5 GPa. Three top most patterns of Fig. \ref{fig6} represent spectra while releasing the pressure. The high-pressure phase is found to be stable down to 11.1 GPa below which the spectrum transforms to a completely different profile with sharp well-defined Raman modes not matching either to any of the two high-pressure phases or to the initial ambient one, establishing that during decompression the system takes a disparate structural transformation pathway to the metastable monoclinic phase as established by our XRD results. Group theory predicts a total of 36 Raman active modes ($\Gamma_{Raman}$ = 18$A_g$ + 18$B_g$) for this pressure-released monoclinic structure among which we have observed 22 modes in the frequency range 30-950 cm$^{-1}$. The highly dissimilar spectral layout of this phase compared to that of the other phases of this system (ambient cubic, high-pressure monoclinic and triclinic) confirms the unique polyhedral configuration of this pressure-released monoclinic structure.

\begin{figure}[h!]
\includegraphics[width=75mm,clip]{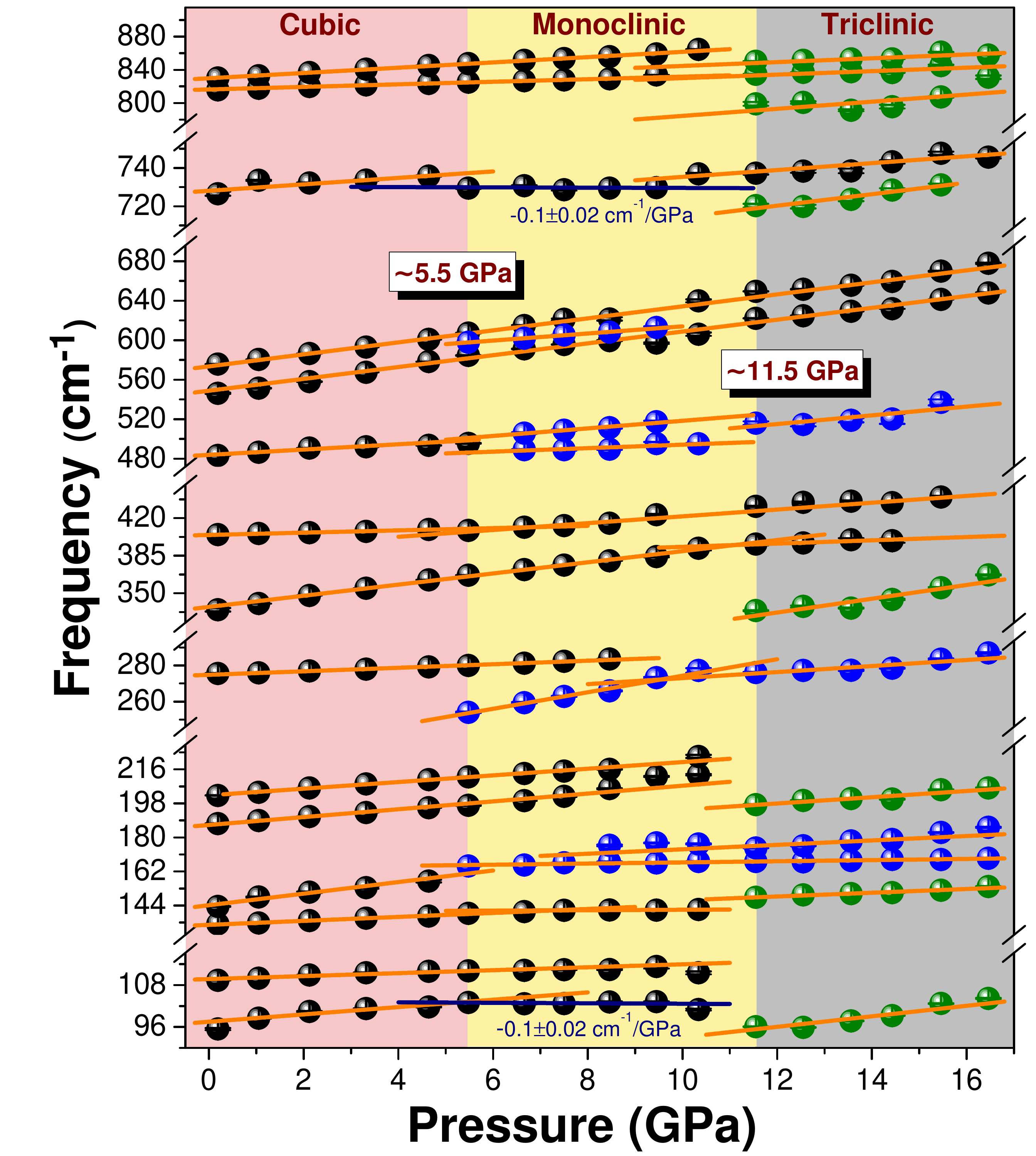}
\caption{\small{Pressure evolution of selected phonon frequencies of Cu$_2$OSeO$_3$. Black solid circles denote phonon modes of ambient cubic phase (shaded pink). New modes appearing at the onsets of the monoclinic (shaded yellow)and triclinic (shaded grey) transitions are denoted by blue and green solid circles, respectively. Solid lines are linear fits of phonon frequencies with pressure.}}
\label{fig7}
\end{figure}

All the Raman modes have been fitted with Lorentzian profile and the mode frequencies are plotted against pressure in Fig. \ref{fig7}. The straight lines represent fitting of mode frequencies using the linear equation $\omega_P = \omega_0 +(\frac{d\omega}{dP})P$. The frequencies of phonon modes, their $\frac{d\omega}{dP}$ values in different phases and the corresponding Gr\"{u}neisen parameters $\gamma_i=\frac{B_0}{\omega}\frac{d\omega}{dP}$ are listed in Table \ref{table5}. All the modes exhibit normal hardening behavior with increasing pressure as expected due to pressure enhancement of stiffness constant due to shrinkage of the unit cell. Fig. \ref{fig7} features the two structural transitions in Cu$_2$OSeO$_3$ at $\sim$ 5 and $\sim$11 GPa with new modes appearing (indicated by blue and green solid circles) and alteration of the slopes of the phonon modes across the transition pressures. The two high-pressure phases are associated with positive slope values for the phonon modes except for the two modes at $\sim$95 and 726 cm$^{-1}$ associated with the initial cubic phase showing mild softening with slightly negative slope value of -0.1 $\pm$ 0.02 cm$^{-1}$/GPa (indicated by navy blue solid lines) in the pressure range 5-10 GPa.

\section{Conclusion}

To summarize, structural and vibrational properties of chiral B20 magnet Cu$_2$OSeO$_3$ has been investigated at room temperature under high pressure using x-ray diffraction and Raman scattering studies. The ambient cubic phase (\emph{P2$_1$3}, \#198) transforms to monoclinic (\emph{P12$_1$1}, \#4) at $\sim$7 GPa and to triclinic (\emph{P1}, \#1) at $\sim$11 GPa both the transitions being first order. The transitions are path-dependent in nature and the system adopts another monoclinic structure (\emph{P12$_1$/c1}, \#14) up on decompressing back to ambient. It would be interesting to calculate magnetic exchange interactions in this pressure-released new monoclinic structure based on our observed structural parameters. The inter-atomic distances of the copper tetrahedra falls off with increasing pressure in the cubic phase and is responsible for the increasing \emph{T$_C$} value as well as the growing pocket size of the skyrmion phase. High pressure Raman studies support the two structural transitions with emergence of new vibrational modes in the spectra and changes in pressure derivative of the phonon frequencies across the transition pressures. Our first-principles calculations  for the ambient cubic phase  (\emph{P2$_1$3}, \#198) reveal that hydrostatic pressure affects spin-spin exchange interactions significantly, and pressure is likely to enhance the size of skyrmion pocket in Cu$_2$OSeO$_3$. These observations should open up future scope for detailed theoretical and experimental studies to unleash the microscopic magnetic configurations of these new high-pressure structures of Cu$_2$OSeO$_3$. 

\section*{Acknowledgments}

We thank Boby Joseph for his support during XRD measurements at the Xpress beamline of Elettra Sincrotrone Trieste~\cite{Lotti}. Financial supports by Department of Science and Technology (Government of India) is also gratefully acknowledged. AKS thanks Nanomission Council and the Year of Science professorship of DST for financial support. SC acknowledges JNCASR for reasearch fellowship. UVW thanks DST, Govt. of India for support through a J.C. Bose National fellowship of SERB.

\section*{}

\section*{Figures and Tables}

\begin{table*}[!ht]
\footnotesize
    \caption{\footnotesize Refined Lattice Parameters for different phases of Cu$_2$OSeO$_3$}
    \begin{center}
    \begin{tabularx}{\textwidth}{X X X X X}
    \arrayrulecolor{black}\hline \hline \\ [-1.5ex]
    \multicolumn{1}{X}{}
    & \multicolumn{1}{X}{\textbf{\hspace{1.5ex} Cubic}}
    & \multicolumn{1}{X}{\textbf{Monoclinic}}
    & \multicolumn{1}{X}{\textbf{\hspace{0.8ex} Triclinic}} 
    & \multicolumn{1}{X}{\textbf{\hspace{0.8ex}Monoclinic}} \\
    { } & \textbf{(1.1 GPa)} & \textbf{(10.5 GPa)} & \textbf{(12.7 GPa)} & \textbf{(3.4 GPa\_R)}\\  [1ex]   	\hline \\ [-1.5ex]
    \textbf{Space Group} & \emph{P2$_1$3} & \emph{P12$_1$1} & \emph{P1} & \emph{P12$_1$/c1}\\[1ex]
    \textbf{a, b, c} \boldsymbol{$({\AA})$} & a=$8.88921$=b=c & a=$8.487775,$ & a=$8.499870,$ & a=$6.926283,$\\
    { } & { } & {b=$9.222242,$} & {b=$9.230770,$} & {b=$5.868629,$}\\
    { } & { } & {c=$8.005783$} & {c=$7.920129$} & {c=$10.561576$}\\[1ex]
    \boldsymbol{$\alpha, \beta, \gamma\;(\degree)$} & $\alpha=\beta=\gamma=90$ & $\alpha=\gamma=90, $ & $\alpha=90.809,$ & $\alpha=\gamma=90,$\\
    { } & { } & $\beta=92.307$ & $\beta=92.809,$ & $\beta=128.595$ \\
    { } & { } & { } & $\gamma=90.825$ & { }\\[1ex]
    \textbf{V/f.u.} \boldsymbol{$({\AA}^3)$} & $87.801$ (z=8) & $78.269$ (z=8) & $77.594$ (z=8) & $83.883$ (z=4)\\[1ex]
    \hline \hline
    \end{tabularx}
    \end{center}
    \label{table1}
\end{table*}

\begin{table}[!ht]
\footnotesize
    \caption{\footnotesize Estimated value of $D$ and $J$ parameters and the distance (d) between respective Cu-atoms.}
    \begin{center}
    \begin{tabularx}{250pt}{>{\hsize=.8\hsize\linewidth=\hsize}X
                                                     >{\hsize=1\hsize\linewidth=\hsize}X
                                                     >{\hsize=1.5\hsize\linewidth=\hsize}X
                                                     >{\hsize=.5\hsize\linewidth=\hsize}X }
    \hline \hline \\ [-1.5ex]
Coupling & J (meV) & D (meV) & d (\AA) \\ [1ex]
\hline \\ [-1.5ex]
1. & -2.12 & (0.53, 0.51, -0.08) & 3.07 \\
2. & 12.46 & (-2.64, -0.6, -2.13) & 3.07 \\
3. & -6.77 & (-0.90, -0.61, -0.38) & 3.27 \\
4. & 2.10 & (0.89, 2.1, -3.17) & 3.35 \\
5. & 2.54 & (-0.12, 0.19, -0.09) & 6.41 \\ [1ex]
\hline \hline

    \end{tabularx}
    \end{center}
    \label{table2}
\end{table}

\begin{table}[!ht]
\footnotesize
    \caption{\footnotesize Pressure dependence of $J$-coupling constants. All J-couplings strengthen with pressure except $J_{3}$.~Super-superexchange interaction $J_{5}$ shows a small increase from 0 to 6 GPa.}
    \begin{center}
    \begin{tabularx}{250pt}{>{\hsize=.8\hsize\linewidth=\hsize}X
                                                     >{\hsize=1\hsize\linewidth=\hsize}X
                                                     >{\hsize=1\hsize\linewidth=\hsize}X
                                                     >{\hsize=1\hsize\linewidth=\hsize}X }
    \hline \hline \\[-1.5ex]
J (meV) & at 0 GPa & at 3 GPa & at 6 GPa \\ [1ex]
\hline \\[-1.5ex]
$J_{1}$  & -2.12 & -2.56 & -2.99 \\
$J_{2}$  & 12.46 & 12.53 & 12.66 \\
$J_{3}$  & -6.77 & -6.60 & -6.57 \\
$J_{4}$  & 2.10 &2.52 & 3.0 \\
$J_{5}$  &2.54 & 2.57 & 2.62 \\ [1ex]
\hline \hline

    \end{tabularx}
    \end{center}
    \label{table3}
\end{table}

\begin{table*}[!ht]
\footnotesize
    \caption{\footnotesize Phonon mode frequencies, their pressure derivatives and the corresponding Gr\"{u}neisen parameters for ambient and high pressure phases of Cu$_2$OSeO$_3$. Non-existence of modes in regions are denoted by $---$.}
    \begin{center}
    \begin{tabularx}{\textwidth}{X X X X X X}
    \arrayrulecolor{black}\hline \hline \\[-1.5ex]
    \multirow{2}{3cm}{\centering\textbf{Phase}} & \multirow{2}{3.2cm}{\centering\boldsymbol{$\omega\;(cm^{-1})$}} & \multicolumn{3}{c}{\boldsymbol{$\frac{d\omega}{dP}\:(cm^{-1}/GPa)$}} & \multirow{2}{4cm}{\centering\boldsymbol{$\gamma_i=\frac{B_0}{\omega}\frac{d\omega}{dP}$}}\\
    {} & {} & \multicolumn{1}{c}{\textbf{I}} & \multicolumn{1}{c}{\textbf{II}} & \multicolumn{1}{c}{\textbf{III}} & {}\\[1ex]
    \hline \\[-1.5ex]
    \multirow{15}{3cm}{\centering{Cubic (I)\\ (B$_0$=74.5 GPa)}} & \multicolumn{1}{c}{95.4$\pm$0.3} & \multicolumn{1}{c}{1.0$\pm$0.2} & \multicolumn{1}{c}{-0.1$\pm$0.02} & \multicolumn{1}{c}{$---$} & \multicolumn{1}{c}{0.8}\\
    {} & \multicolumn{1}{c}{109.3$\pm$0.05} & \multicolumn{2}{c}{0.4$\pm$0.06} & \multicolumn{1}{c}{$---$} & \multicolumn{1}{c}{0.3}\\
    {} & \multicolumn{1}{c}{134.2$\pm$0.05} & \multicolumn{1}{c}{1.0$\pm$0.05} & \multicolumn{1}{c}{0.1$\pm$0.04} & \multicolumn{1}{c}{$---$} & \multicolumn{1}{c}{0.5}\\
    {} & \multicolumn{1}{c}{143.5$\pm$0.3} & \multicolumn{1}{c}{3.1$\pm$0.5} & \multicolumn{1}{c}{$---$} & \multicolumn{1}{c}{$---$} & \multicolumn{1}{c}{1.6}\\
    {} & \multicolumn{1}{c}{187.3$\pm$0.05} & \multicolumn{2}{c}{2.1$\pm$0.1} & \multicolumn{1}{c}{$---$} & \multicolumn{1}{c}{0.8}\\
    {} & \multicolumn{1}{c}{202.2$\pm$0.06} & \multicolumn{2}{c}{1.7$\pm$0.05} & \multicolumn{1}{c}{$---$} & \multicolumn{1}{c}{0.6}\\
    {} & \multicolumn{1}{c}{275.6$\pm$0.1} & \multicolumn{2}{c}{1.0$\pm$0.03} & \multicolumn{1}{c}{$---$} & \multicolumn{1}{c}{0.3}\\
    {} & \multicolumn{1}{c}{334.5$\pm$0.2} & \multicolumn{2}{c}{5.2$\pm$0.1} & \multicolumn{1}{c}{1.5$\pm$0.5} & \multicolumn{1}{c}{1.1}\\
    {} & \multicolumn{1}{c}{404.6$\pm$0.6} & \multicolumn{1}{c}{1.0$\pm$0.1} & \multicolumn{2}{c}{3.2$\pm$0.3} & \multicolumn{1}{c}{0.2}\\
    {} & \multicolumn{1}{c}{483.4$\pm$0.1} & \multicolumn{1}{c}{2.7$\pm$0.5} & \multicolumn{1}{c}{$---$} & \multicolumn{1}{c}{$---$} & \multicolumn{1}{c}{0.4}\\
    {} & \multicolumn{1}{c}{546.2$\pm$0.5} & \multicolumn{3}{c}{6.0$\pm$0.1} & \multicolumn{1}{c}{0.8}\\
    {} & \multicolumn{1}{c}{575.8$\pm$0.1} & \multicolumn{3}{c}{6.1$\pm$0.2} & \multicolumn{1}{c}{0.8}\\
    {} & \multicolumn{1}{c}{726.5$\pm$1.0} &  \multicolumn{1}{c}{1.7$\pm$0.3} & \multicolumn{1}{c}{-0.1$\pm$0.02} & \multicolumn{1}{c}{1.6$\pm$0.3} & \multicolumn{1}{c}{0.2}\\
    {} & \multicolumn{1}{c}{815.8$\pm$0.04} & \multicolumn{2}{c}{1.5$\pm$0.04} & \multicolumn{1}{c}{$---$} & \multicolumn{1}{c}{0.1}\\
    {} & \multicolumn{1}{c}{830.1$\pm$0.02} & \multicolumn{2}{c}{3.2$\pm$0.05} & \multicolumn{1}{c}{$---$} & \multicolumn{1}{c}{0.3}\\[1ex]
    \hline \\[-1.5ex]
    \multirow{6}{3cm}{\centering{Monoclinic (II)\\ (B$_0$=161.1 GPa)}} & \multicolumn{1}{c}{164.9$\pm$0.2} & \multicolumn{1}{c}{$---$} & \multicolumn{2}{c}{0.3$\pm$0.04} & \multicolumn{1}{c}{0.3}\\
    {} & \multicolumn{1}{c}{175.8$\pm$0.4} & \multicolumn{1}{c}{$---$} & \multicolumn{2}{c}{1.2$\pm$0.3} & \multicolumn{1}{c}{1.1}\\
    {} & \multicolumn{1}{c}{253.9$\pm$0.3} & \multicolumn{1}{c}{$---$} & \multicolumn{1}{c}{4.6$\pm$0.3} & \multicolumn{1}{c}{1.7$\pm$0.3} & \multicolumn{1}{c}{2.9}\\
     {} & \multicolumn{1}{c}{489.1$\pm$0.6} & \multicolumn{1}{c}{$---$} & \multicolumn{1}{c}{1.7$\pm$0.4} & \multicolumn{1}{c}{$---$} & \multicolumn{1}{c}{0.5}\\
     {} & \multicolumn{1}{c}{506.0$\pm$0.7} & \multicolumn{1}{c}{$---$} & \multicolumn{1}{c}{3.8$\pm$0.5} & \multicolumn{1}{c}{$---$} & \multicolumn{1}{c}{1.2}\\
     {} & \multicolumn{1}{c}{597.9$\pm$0.2} & \multicolumn{1}{c}{$---$} & \multicolumn{1}{c}{3.6$\pm$0.1} & \multicolumn{1}{c}{4.4$\pm$1.7} & \multicolumn{1}{c}{1.0}\\[1ex]
     \hline \\[-1.5ex]
     \multirow{8}{3cm}{\centering{Triclinic (III)\\ (B$_0$=183.3 GPa)}} & \multicolumn{1}{c}{96.1$\pm$0.06} & \multicolumn{1}{c}{$---$} & \multicolumn{1}{c}{$---$} &  \multicolumn{1}{c}{1.5$\pm$0.3} & \multicolumn{1}{c}{2.9}\\
     {} & \multicolumn{1}{c}{148.3$\pm$0.06} & \multicolumn{1}{c}{$---$} & \multicolumn{1}{c}{$---$} &  \multicolumn{1}{c}{1.0$\pm$0.1} & \multicolumn{1}{c}{1.2}\\
     {} & \multicolumn{1}{c}{197.4$\pm$0.2} & \multicolumn{1}{c}{$---$} & \multicolumn{1}{c}{$---$} &  \multicolumn{1}{c}{1.6$\pm$0.3} & \multicolumn{1}{c}{1.5}\\
     {} & \multicolumn{1}{c}{333.7$\pm$0.5} & \multicolumn{1}{c}{$---$} & \multicolumn{1}{c}{$---$} &  \multicolumn{1}{c}{6.4$\pm$1.5} & \multicolumn{1}{c}{3.5}\\
     {} & \multicolumn{1}{c}{720.3$\pm$0.9} & \multicolumn{1}{c}{$---$} & \multicolumn{1}{c}{$---$} &  \multicolumn{1}{c}{3.0$\pm$0.3} & \multicolumn{1}{c}{0.8}\\
     {} & \multicolumn{1}{c}{799.0$\pm$2.3} & \multicolumn{1}{c}{$---$} & \multicolumn{1}{c}{$---$} &  \multicolumn{1}{c}{1.8$\pm$0.4} & \multicolumn{1}{c}{0.4}\\
     {} & \multicolumn{1}{c}{834.6$\pm$0.2} & \multicolumn{1}{c}{$---$} & \multicolumn{1}{c}{$---$} &  \multicolumn{1}{c}{2.1$\pm$1.0} & \multicolumn{1}{c}{0.5}\\
     {} & \multicolumn{1}{c}{850.2$\pm$0.1} & \multicolumn{1}{c}{$---$} & \multicolumn{1}{c}{$---$} &  \multicolumn{1}{c}{2.3$\pm$0.4} & \multicolumn{1}{c}{0.5}\\[1ex]
     \hline \hline
    \end{tabularx}
    \end{center}
    \label{table5}
\end{table*}

\end{document}